\documentstyle[bo99,epsfig]{article}

\def\oiii{[{\sc O\, iii}]}

\def\feka{Fe K$\alpha$}
\def\ltsima{$\; \buildrel < \over \sim \;$}
\def\simlt{\lower.5ex\hbox{\ltsima}} 
\def\gtsima{$\; \buildrel > \over \sim \;$}
\def\simgt{\lower.5ex\hbox{\gtsima}} 

\title{Radio-loud AGNs: The X-Ray Perspective} 
\author{Rita M. Sambruna and Michael Eracleous} 
\affil{Pennsylvania State University,
Dept. of Astron. \& Astrophys., 525 Davey Lab, University Park 16802
(emails: rms@astro.psu.edu, mce@astro.psu.edu)}

\begin{document}
\def\PsfigVersion{1.9}
\ifx\undefined\psfig\else \fi

%

\let\LaTeXAtSign=\@
\let\@=\relax
\edef\psfigRestoreAt{\catcode`\@=\number\catcode`@\relax}
\catcode`\@=11\relax
\newwrite\@unused
\def\ps@typeout#1{{\let\protect\string\immediate\write\@unused{#1}}}
\ps@typeout{psfig/tex \PsfigVersion}


\def\figurepath{./}
\def\psfigurepath#1{\edef\figurepath{#1}}

%
%
\def\@nnil{\@nil}
\def\@empty{}
\def\@psdonoop#1\@@#2#3{}
\def\@psdo#1:=#2\do#3{\edef\@psdotmp{#2}\ifx\@psdotmp\@empty \else
    \expandafter\@psdoloop#2,\@nil,\@nil\@@#1{#3}\fi}
\def\@psdoloop#1,#2,#3\@@#4#5{\def#4{#1}\ifx #4\@nnil \else
       #5\def#4{#2}\ifx #4\@nnil \else#5\@ipsdoloop #3\@@#4{#5}\fi\fi}
\def\@ipsdoloop#1,#2\@@#3#4{\def#3{#1}\ifx #3\@nnil 
       \let\@nextwhile=\@psdonoop \else
      #4\relax\let\@nextwhile=\@ipsdoloop\fi\@nextwhile#2\@@#3{#4}}
\def\@tpsdo#1:=#2\do#3{\xdef\@psdotmp{#2}\ifx\@psdotmp\@empty \else
    \@tpsdoloop#2\@nil\@nil\@@#1{#3}\fi}
\def\@tpsdoloop#1#2\@@#3#4{\def#3{#1}\ifx #3\@nnil 
       \let\@nextwhile=\@psdonoop \else
      #4\relax\let\@nextwhile=\@tpsdoloop\fi\@nextwhile#2\@@#3{#4}}
%
\ifx\undefined\fbox
\newdimen\fboxrule
\newdimen\fboxsep
\newdimen\ps@tempdima
\newbox\ps@tempboxa
\fboxsep = 3pt
\fboxrule = .4pt
\long\def\fbox#1{\leavevmode\setbox\ps@tempboxa\hbox{#1}\ps@tempdima\fboxrule
    \advance\ps@tempdima \fboxsep \advance\ps@tempdima \dp\ps@tempboxa
   \hbox{\lower \ps@tempdima\hbox
  {\vbox{\hrule height \fboxrule
          \hbox{\vrule width \fboxrule \hskip\fboxsep
          \vbox{\vskip\fboxsep \box\ps@tempboxa\vskip\fboxsep}\hskip 
                 \fboxsep\vrule width \fboxrule}
                 \hrule height \fboxrule}}}}
\fi
%
%
\newread\ps@stream
\newif\ifnot@eof       
\newif\if@noisy        
\newif\if@atend        
\newif\if@psfile       
%
%
{\catcode`\%=12\global\gdef\epsf@start{
\def\epsf@PS{PS}
\def\epsf@getbb#1{%
%
%
\openin\ps@stream=#1
\ifeof\ps@stream\ps@typeout{Error, File #1 not found}\else
%
%
   {\not@eoftrue \chardef\other=12
    \def\do##1{\catcode`##1=\other}\dospecials \catcode`\ =10
    \loop
       \if@psfile
	  \read\ps@stream to \epsf@fileline
       \else{
	  \obeyspaces
          \read\ps@stream to \epsf@tmp\global\let\epsf@fileline\epsf@tmp}
       \fi
       \ifeof\ps@stream\not@eoffalse\else
%
%
       \if@psfile\else
       \expandafter\epsf@test\epsf@fileline:. \\%
       \fi
%
%
          \expandafter\epsf@aux\epsf@fileline:. \\%
       \fi
   \ifnot@eof\repeat
   }\closein\ps@stream\fi}%
%
%
\long\def\epsf@test#1#2#3:#4\\{\def\epsf@testit{#1#2}
			\ifx\epsf@testit\epsf@start\else
\ps@typeout{Warning! File does not start with `\epsf@start'.  It may not be a PostScript file.}
			\fi
			\@psfiletrue} 
%
%
{\catcode`\%=12\global\let\epsf@percent=
%
%
%
\long\def\epsf@aux#1#2:#3\\{\ifx#1\epsf@percent
   \def\epsf@testit{#2}\ifx\epsf@testit\epsf@bblit
	\@atendfalse
        \epsf@atend #3 . \\%
	\if@atend	
	   \if@verbose{
		\ps@typeout{psfig: found `(atend)'; continuing search}
	   }\fi
        \else
        \epsf@grab #3 . . . \\%
        \not@eoffalse
        \global\no@bbfalse
        \fi
   \fi\fi}%
%
%
\def\epsf@grab #1 #2 #3 #4 #5\\{%
   \global\def\epsf@llx{#1}\ifx\epsf@llx\empty
      \epsf@grab #2 #3 #4 #5 .\\\else
   \global\def\epsf@lly{#2}%
   \global\def\epsf@urx{#3}\global\def\epsf@ury{#4}\fi}%
%
%
\def\epsf@atendlit{(atend)} 
\def\epsf@atend #1 #2 #3\\{%
   \def\epsf@tmp{#1}\ifx\epsf@tmp\empty
      \epsf@atend #2 #3 .\\\else
   \ifx\epsf@tmp\epsf@atendlit\@atendtrue\fi\fi}


\chardef\psletter = 11 
\chardef\other = 12

\newif \ifdebug 
\newif\ifc@mpute 
\c@mputetrue 

\let\then = \relax
\def\r@dian{pt }
\let\r@dians = \r@dian
\let\dimensionless@nit = \r@dian
\let\dimensionless@nits = \dimensionless@nit
\def\internal@nit{sp }
\let\internal@nits = \internal@nit
\newif\ifstillc@nverging
\def \Mess@ge #1{\ifdebug \then \message {#1} \fi}

{ 
	\catcode `\@ = \psletter
	\gdef \nodimen {\expandafter \n@dimen \the \dimen}
	\gdef \term #1 #2 #3%
	       {\edef \t@ {\the #1}
		\edef \t@@ {\expandafter \n@dimen \the #2\r@dian}%
		\t@rm {\t@} {\t@@} {#3}%
	       }
	\gdef \t@rm #1 #2 #3%
	       {{%
		\count 0 = 0
		\dimen 0 = 1 \dimensionless@nit
		\dimen 2 = #2\relax
		\Mess@ge {Calculating term #1 of \nodimen 2}%
		\loop
		\ifnum	\count 0 < #1
		\then	\advance \count 0 by 1
			\Mess@ge {Iteration \the \count 0 \space}%
			\Multiply \dimen 0 by {\dimen 2}%
			\Mess@ge {After multiplication, term = \nodimen 0}%
			\Divide \dimen 0 by {\count 0}%
			\Mess@ge {After division, term = \nodimen 0}%
		\repeat
		\Mess@ge {Final value for term #1 of 
				\nodimen 2 \space is \nodimen 0}%
		\xdef \Term {#3 = \nodimen 0 \r@dians}%
		\aftergroup \Term
	       }}
	\catcode `\p = \other
	\catcode `\t = \other
	\gdef \n@dimen #1pt{#1} 
}

\def \Divide #1by #2{\divide #1 by #2} 

\def \Multiply #1by #2
       {{
	\count 0 = #1\relax
	\count 2 = #2\relax
	\count 4 = 65536
	\Mess@ge {Before scaling, count 0 = \the \count 0 \space and
			count 2 = \the \count 2}%
	\ifnum	\count 0 > 32767 
	\then	\divide \count 0 by 4
		\divide \count 4 by 4
	\else	\ifnum	\count 0 < -32767
		\then	\divide \count 0 by 4
			\divide \count 4 by 4
		\else
		\fi
	\fi
	\ifnum	\count 2 > 32767 
	\then	\divide \count 2 by 4
		\divide \count 4 by 4
	\else	\ifnum	\count 2 < -32767
		\then	\divide \count 2 by 4
			\divide \count 4 by 4
		\else
		\fi
	\fi
	\multiply \count 0 by \count 2
	\divide \count 0 by \count 4
	\xdef \product {#1 = \the \count 0 \internal@nits}%
	\aftergroup \product
       }}

\def\r@duce{\ifdim\dimen0 > 90\r@dian \then   
		\multiply\dimen0 by -1
		\advance\dimen0 by 180\r@dian
		\r@duce
	    \else \ifdim\dimen0 < -90\r@dian \then  
		\advance\dimen0 by 360\r@dian
		\r@duce
		\fi
	    \fi}

\def\Sine#1%
       {{%
	\dimen 0 = #1 \r@dian
	\r@duce
	\ifdim\dimen0 = -90\r@dian \then
	   \dimen4 = -1\r@dian
	   \c@mputefalse
	\fi
	\ifdim\dimen0 = 90\r@dian \then
	   \dimen4 = 1\r@dian
	   \c@mputefalse
	\fi
	\ifdim\dimen0 = 0\r@dian \then
	   \dimen4 = 0\r@dian
	   \c@mputefalse
	\fi
	\ifc@mpute \then
		\divide\dimen0 by 180
		\dimen0=3.141592654\dimen0
		\dimen 2 = 3.1415926535897963\r@dian 
		\divide\dimen 2 by 2 
		\Mess@ge {Sin: calculating Sin of \nodimen 0}%
		\count 0 = 1 
		\dimen 2 = 1 \r@dian 
		\dimen 4 = 0 \r@dian 
		\loop
			\ifnum	\dimen 2 = 0 
			\then	\stillc@nvergingfalse 
			\else	\stillc@nvergingtrue
			\fi
			\ifstillc@nverging 
			\then	\term {\count 0} {\dimen 0} {\dimen 2}%
				\advance \count 0 by 2
				\count 2 = \count 0
				\divide \count 2 by 2
				\ifodd	\count 2 
				\then	\advance \dimen 4 by \dimen 2
				\else	\advance \dimen 4 by -\dimen 2
				\fi
		\repeat
	\fi		
			\xdef \sine {\nodimen 4}%
       }}

\def\Cosine#1{\ifx\sine\UnDefined\edef\Savesine{\relax}\else
		             \edef\Savesine{\sine}\fi
	{\dimen0=#1\r@dian\advance\dimen0 by 90\r@dian
	 \Sine{\nodimen 0}
	 \xdef\cosine{\sine}
	 \xdef\sine{\Savesine}}}	      

\def\psdraft{
	\def\@psdraft{0}
}
\def\psfull{
	\def\@psdraft{100}
}

\psfull

\newif\if@scalefirst
\def\psscalefirst{\@scalefirsttrue}
\def\psrotatefirst{\@scalefirstfalse}
\psrotatefirst

\newif\if@draftbox
\def\psnodraftbox{
	\@draftboxfalse
}
\def\psdraftbox{
	\@draftboxtrue
}
\@draftboxtrue

\newif\if@prologfile
\newif\if@postlogfile
\def\pssilent{
	\@noisyfalse
}
\def\psnoisy{
	\@noisytrue
}
\psnoisy
\newif\if@bbllx
\newif\if@bblly
\newif\if@bburx
\newif\if@bbury
\newif\if@height
\newif\if@width
\newif\if@rheight
\newif\if@rwidth
\newif\if@angle
\newif\if@clip
\newif\if@verbose
\def\@p@@sclip#1{\@cliptrue}

\newif\if@decmpr


\def\@p@@sfigure#1{\def\@p@sfile{null}\def\@p@sbbfile{null}
	        \openin1=#1.bb
		\ifeof1\closein1
	        	\openin1=\figurepath#1.bb
			\ifeof1\closein1
			        \openin1=#1
				\ifeof1\closein1%
				       \openin1=\figurepath#1
					\ifeof1
					   \ps@typeout{Error, File #1 not found}
						\if@bbllx\if@bblly
				   		\if@bburx\if@bbury
			      				\def\@p@sfile{#1}%
			      				\def\@p@sbbfile{#1}%
							\@decmprfalse
				  	   	\fi\fi\fi\fi
					\else\closein1
				    		\def\@p@sfile{\figurepath#1}%
				    		\def\@p@sbbfile{\figurepath#1}%
						\@decmprfalse
	                       		\fi%
			 	\else\closein1%
					\def\@p@sfile{#1}
					\def\@p@sbbfile{#1}
					\@decmprfalse
			 	\fi
			\else
				\def\@p@sfile{\figurepath#1}
				\def\@p@sbbfile{\figurepath#1.bb}
				\@decmprtrue
			\fi
		\else
			\def\@p@sfile{#1}
			\def\@p@sbbfile{#1.bb}
			\@decmprtrue
		\fi}

\def\@p@@sfile#1{\@p@@sfigure{#1}}

\def\@p@@sbbllx#1{
		\@bbllxtrue
		\dimen100=#1
		\edef\@p@sbbllx{\number\dimen100}
}
\def\@p@@sbblly#1{
		\@bbllytrue
		\dimen100=#1
		\edef\@p@sbblly{\number\dimen100}
}
\def\@p@@sbburx#1{
		\@bburxtrue
		\dimen100=#1
		\edef\@p@sbburx{\number\dimen100}
}
\def\@p@@sbbury#1{
		\@bburytrue
		\dimen100=#1
		\edef\@p@sbbury{\number\dimen100}
}
\def\@p@@sheight#1{
		\@heighttrue
		\dimen100=#1
   		\edef\@p@sheight{\number\dimen100}
}
\def\@p@@swidth#1{
		\@widthtrue
		\dimen100=#1
		\edef\@p@swidth{\number\dimen100}
}
\def\@p@@srheight#1{
		\@rheighttrue
		\dimen100=#1
		\edef\@p@srheight{\number\dimen100}
}
\def\@p@@srwidth#1{
		\@rwidthtrue
		\dimen100=#1
		\edef\@p@srwidth{\number\dimen100}
}
\def\@p@@sangle#1{
		\@angletrue
		\edef\@p@sangle{#1} 
}
\def\@p@@ssilent#1{ 
		\@verbosefalse
}
\def\@p@@sprolog#1{\@prologfiletrue\def\@prologfileval{#1}}
\def\@p@@spostlog#1{\@postlogfiletrue\def\@postlogfileval{#1}}
\def\@cs@name#1{\csname #1\endcsname}
\def\@setparms#1=#2,{\@cs@name{@p@@s#1}{#2}}
%
%
\def\ps@init@parms{
		\@bbllxfalse \@bbllyfalse
		\@bburxfalse \@bburyfalse
		\@heightfalse \@widthfalse
		\@rheightfalse \@rwidthfalse
		\def\@p@sbbllx{}\def\@p@sbblly{}
		\def\@p@sbburx{}\def\@p@sbbury{}
		\def\@p@sheight{}\def\@p@swidth{}
		\def\@p@srheight{}\def\@p@srwidth{}
		\def\@p@sangle{0}
		\def\@p@sfile{} \def\@p@sbbfile{}
		\def\@p@scost{10}
		\def\@sc{}
		\@prologfilefalse
		\@postlogfilefalse
		\@clipfalse
		\if@noisy
			\@verbosetrue
		\else
			\@verbosefalse
		\fi
}
%
%
\def\parse@ps@parms#1{
	 	\@psdo\@psfiga:=#1\do
		   {\expandafter\@setparms\@psfiga,}}
%
%
\newif\ifno@bb
\def\bb@missing{
	\if@verbose{
		\ps@typeout{psfig: searching \@p@sbbfile \space  for bounding box}
	}\fi
	\no@bbtrue
	\epsf@getbb{\@p@sbbfile}
        \ifno@bb \else \bb@cull\epsf@llx\epsf@lly\epsf@urx\epsf@ury\fi
}	
\def\bb@cull#1#2#3#4{
	\dimen100=#1 bp\edef\@p@sbbllx{\number\dimen100}
	\dimen100=#2 bp\edef\@p@sbblly{\number\dimen100}
	\dimen100=#3 bp\edef\@p@sbburx{\number\dimen100}
	\dimen100=#4 bp\edef\@p@sbbury{\number\dimen100}
	\no@bbfalse
}
\newdimen\p@intvaluex
\newdimen\p@intvaluey
\def\rotate@#1#2{{\dimen0=#1 sp\dimen1=#2 sp
		  \global\p@intvaluex=\cosine\dimen0
		  \dimen3=\sine\dimen1
		  \global\advance\p@intvaluex by -\dimen3
		  \global\p@intvaluey=\sine\dimen0
		  \dimen3=\cosine\dimen1
		  \global\advance\p@intvaluey by \dimen3
		  }}
\def\compute@bb{
		\no@bbfalse
		\if@bbllx \else \no@bbtrue \fi
		\if@bblly \else \no@bbtrue \fi
		\if@bburx \else \no@bbtrue \fi
		\if@bbury \else \no@bbtrue \fi
		\ifno@bb \bb@missing \fi
		\ifno@bb \ps@typeout{FATAL ERROR: no bb supplied or found}
			\no-bb-error
		\fi
		%
%
		\count203=\@p@sbburx
		\count204=\@p@sbbury
		\advance\count203 by -\@p@sbbllx
		\advance\count204 by -\@p@sbblly
		\edef\ps@bbw{\number\count203}
		\edef\ps@bbh{\number\count204}
		\if@angle 
			\Sine{\@p@sangle}\Cosine{\@p@sangle}
	        	{\dimen100=\maxdimen\xdef\r@p@sbbllx{\number\dimen100}
					    \xdef\r@p@sbblly{\number\dimen100}
			                    \xdef\r@p@sbburx{-\number\dimen100}
					    \xdef\r@p@sbbury{-\number\dimen100}}
%
                        \def\minmaxtest{
			   \ifnum\number\p@intvaluex<\r@p@sbbllx
			      \xdef\r@p@sbbllx{\number\p@intvaluex}\fi
			   \ifnum\number\p@intvaluex>\r@p@sbburx
			      \xdef\r@p@sbburx{\number\p@intvaluex}\fi
			   \ifnum\number\p@intvaluey<\r@p@sbblly
			      \xdef\r@p@sbblly{\number\p@intvaluey}\fi
			   \ifnum\number\p@intvaluey>\r@p@sbbury
			      \xdef\r@p@sbbury{\number\p@intvaluey}\fi
			   }
			\rotate@{\@p@sbbllx}{\@p@sbblly}
			\minmaxtest
			\rotate@{\@p@sbbllx}{\@p@sbbury}
			\minmaxtest
			\rotate@{\@p@sbburx}{\@p@sbblly}
			\minmaxtest
			\rotate@{\@p@sbburx}{\@p@sbbury}
			\minmaxtest
			\edef\@p@sbbllx{\r@p@sbbllx}\edef\@p@sbblly{\r@p@sbblly}
			\edef\@p@sbburx{\r@p@sbburx}\edef\@p@sbbury{\r@p@sbbury}
		\fi
		\count203=\@p@sbburx
		\count204=\@p@sbbury
		\advance\count203 by -\@p@sbbllx
		\advance\count204 by -\@p@sbblly
		\edef\@bbw{\number\count203}
		\edef\@bbh{\number\count204}
}
%
%
\def\in@hundreds#1#2#3{\count240=#2 \count241=#3
		     \count100=\count240	
		     \divide\count100 by \count241
		     \count101=\count100
		     \multiply\count101 by \count241
		     \advance\count240 by -\count101
		     \multiply\count240 by 10
		     \count101=\count240	
		     \divide\count101 by \count241
		     \count102=\count101
		     \multiply\count102 by \count241
		     \advance\count240 by -\count102
		     \multiply\count240 by 10
		     \count102=\count240	
		     \divide\count102 by \count241
		     \count200=#1\count205=0
		     \count201=\count200
			\multiply\count201 by \count100
		 	\advance\count205 by \count201
		     \count201=\count200
			\divide\count201 by 10
			\multiply\count201 by \count101
			\advance\count205 by \count201
		     \count201=\count200
			\divide\count201 by 100
			\multiply\count201 by \count102
			\advance\count205 by \count201
		     \edef\@result{\number\count205}
}
\def\compute@wfromh{
		\in@hundreds{\@p@sheight}{\@bbw}{\@bbh}
		\edef\@p@swidth{\@result}
}
\def\compute@hfromw{
	        \in@hundreds{\@p@swidth}{\@bbh}{\@bbw}
		\edef\@p@sheight{\@result}
}
\def\compute@handw{
		\if@height 
			\if@width
			\else
				\compute@wfromh
			\fi
		\else 
			\if@width
				\compute@hfromw
			\else
				\edef\@p@sheight{\@bbh}
				\edef\@p@swidth{\@bbw}
			\fi
		\fi
}
\def\compute@resv{
		\if@rheight \else \edef\@p@srheight{\@p@sheight} \fi
		\if@rwidth \else \edef\@p@srwidth{\@p@swidth} \fi
}
%
\def\compute@sizes{
	\compute@bb
	\if@scalefirst\if@angle
	\if@width
	   \in@hundreds{\@p@swidth}{\@bbw}{\ps@bbw}
	   \edef\@p@swidth{\@result}
	\fi
	\if@height
	   \in@hundreds{\@p@sheight}{\@bbh}{\ps@bbh}
	   \edef\@p@sheight{\@result}
	\fi
	\fi\fi
	\compute@handw
	\compute@resv}

%
%
\def\psfig#1{\vbox {
	%
	\ps@init@parms
	\parse@ps@parms{#1}
	\compute@sizes
	\ifnum\@p@scost<\@psdraft{
		\special{ps::[begin] 	\@p@swidth \space \@p@sheight \space
				\@p@sbbllx \space \@p@sbblly \space
				\@p@sbburx \space \@p@sbbury \space
				startTexFig \space }
		\if@angle
			\special {ps:: \@p@sangle \space rotate \space} 
		\fi
		\if@clip{
			\if@verbose{
				\ps@typeout{(clip)}
			}\fi
			\special{ps:: doclip \space }
		}\fi
		\if@prologfile
		    \special{ps: plotfile \@prologfileval \space } \fi
		\if@decmpr{
			\if@verbose{
				\ps@typeout{psfig: including \@p@sfile.Z \space }
			}\fi
			\special{ps: plotfile "`zcat \@p@sfile.Z" \space }
		}\else{
			\if@verbose{
				\ps@typeout{psfig: including \@p@sfile \space }
			}\fi
			\special{ps: plotfile \@p@sfile \space }
		}\fi
		\if@postlogfile
		    \special{ps: plotfile \@postlogfileval \space } \fi
		\special{ps::[end] endTexFig \space }
		\vbox to \@p@srheight sp{
			\hbox to \@p@srwidth sp{
				\hss
			}
		\vss
		}
	}\else{
		\if@draftbox{		
			\hbox{\frame{\vbox to \@p@srheight sp{
			\vss
			\hbox to \@p@srwidth sp{ \hss \@p@sfile \hss }
			\vss
			}}}
		}\else{
			\vbox to \@p@srheight sp{
			\vss
			\hbox to \@p@srwidth sp{\hss}
			\vss
			}
		}\fi

	}\fi
}}
\psfigRestoreAt
\let\@=\LaTeXAtSign

\maketitle

\begin{abstract}

The X-ray emission of radio-loud (RL) AGNs is a powerful tool for
probing the structure of the accretion flow in these objects. We
review recent spectral and variability studies of RL AGNs, which show
that these systems have systematically different X-ray properties than
their radio-quiet (RQ) counterparts. Specifically, RL AGNs have weaker
and narrower \feka\ lines and weaker Compton reflection components
above 10 keV. The nuclear continuum of RL AGNs in the 2--10 keV band
is well described by a power law with photon indices $\sim$ 1.8,
similar to RQ AGNs of comparable X-ray luminosity.  RL AGNs have
little or no flux variability on short time scales (\ltsima 0.5 days);
however, flux and spectral variations are observed on time scales of
weeks in two well-monitored objects, 3C~390.3 and 3C~120. These
properties strongly suggest that the central engines of the two AGNs
classes are different. We discuss the implications of these
observational results, in particular the possibility that the central
engines of RL AGNs are harbor an ion torus (also known as an
Advection-Dominated Accretion Flow or ADAF). 
We show that a beamed component from the jet is unlikely
in the majority of sources. 
Moreover, the X-ray data provide evidence that the
circumnuclear environs of RL and RQ AGNs also differ: large amounts of
cold gas are detected in BLRGs and QSRs, contrary to Seyfert galaxies
of similar X-ray luminosity where an ionized absorber seems to be the
norm. The role of future X-ray missions in advancing our understanding
of the central engines of RL AGNs is briefly highlighted.


\end{abstract}

\section{The X-ray Advantage}

The spectra and variability properties of AGNs are a diagnostic of the
conditions of the matter in the inner parts of the accretion flow.
Recent studies of X-ray-bright, radio-quiet (RQ) Seyfert galaxies have
provided evidence that their X-ray emission is complex. At high
energies, the most prominent features are the \feka\ emission line
between 6 and 7~keV and the Compton reflection hump at \ltsima 10 keV,
originating from a cold reprocessor near the central black hole.  At
soft X-rays, many Seyferts exhibit absorption from partially ionized
gas with column densities $N_{\rm W} \sim
10^{21}$--$10^{24}$~cm$^{-2}$ along the line of sight. The rapid flux
variations of Seyferts galaxies on timescales of hours and even
minutes (e.g., Edelson 2000) indicate that the X-ray source is compact
and located very close to the central black hole.

In contrast, the X-ray spectra and variability of radio-loud (RL)
AGNs\footnote[1]{Here we will exclude blazars, as their continuum is
entirely dominated by the the beamed jet emission.} are not as well
known. Past X-ray studies with {\it Einstein, EXOSAT} and {\it GINGA}
showed that RL AGNs had systematically harder (i.e, flatter) X-ray
continua than their radio-quiet counterparts (Wilkes \& Elvis 1987;
Shastri et al. 1993; Lawson \& Turner 1997). These studies, however,
were plagued by the low sensitivity and resolution of the instruments,
in addition to the heterogeneity of the samples.

Fundamental progress was recently achieved with the advent of {\it
ASCA} and {\it RXTE} (for a review of the {\it BeppoSAX} results, see
Grandi 2000). The wide-band coverage of these instruments, combined to
their higher sensitivity and/or resolution compared to older
instruments, enable us to disentangle the various spectral components
and study flux and spectral variability. We started a program of
systematic study of the X-ray properties of RL AGNs, with the ultimate
goal of elucidating the structure of the accretion flow in these
systems and comparing it to RQ AGNs. Here we review the results of our
{\it ASCA} and {\it RXTE} surveys (Sambruna, Eracleous, \& Mushotzky
1999; Eracleous, Sambruna, \& Mushotzky 1999; also reporting
references to previous works), and present new observations of
selected objects. 

\section{The {\it ASCA} and {\it RXTE} Databases for RL AGNs}

In the following we concentrate on AGNs with lobe-dominated radio
morphologies, which we define as those objects with either a 5 GHz
radio power of $P_{\rm \,5\,GHz} > 10^{25}~{\rm W~Hz^{-1}}$ or with
rest-frame {5~GHz-to-4400~\AA} flux-density ratios of ${\cal R}_{\rm
ro} > 10$, following Kellermann et al. (1994). There are 39
objects in the {\it ASCA} archive up to September 1998 which satisfy
these criteria, of which 9 are Broad Line Radio Galaxies (BLRGs), 6
are Quasars (QSRs), 12 are Narrow Line Radio Galaxies (NLRGs), and 11
are Radio Galaxies (RGs). This subdivision depends
on their optical spectral properties, namely, the presence of broad,
permitted lines in their optical spectra and the luminosity of the
\oiii~$\lambda5007$ line. Clearly the sample is not statistical or
complete, and most likely is biased toward the brightest sources of
each type. 

Because of its high background rate, only the brightest sources,
typically BLRGs, were observed by {\it RXTE}. Our sample (Eracleous et
al. 1999) includes four BLRGs, namely 3C~120, 3C~111, 3C~382, and
Pictor~A, observed for typically $\sim$ 40 ks during AO1 and
AO2. These have fluxes F$_{2-10~keV} \sim 1-4 \times 10^{-11}$ erg
cm$^{-2}$ s$^{-1}$, ensuring that their spectra could be measured
adequately with both the PCA and HEXTE. The NLRG Centaurus A was also
observed with {\it RXTE}; the results are presented by Rothschild et
al. (1999). 

The BLRGs 3C~390.3 and 3C~120 were intensively monitored with {\it
RXTE} in 1997 May and January for 134 and 150 ks, respectively (P.I.:
R. Remillard), as part of multiwavelength campaigns. Here we present
the preliminary results of our analysis of the archival data. We also
discuss simultaneous {\it ASCA} 100 ks and {\it RXTE} 60 ks
observations of 3C~382 obtained by us in 1999 March.


\section{The Nuclear X-ray Emission of RL AGNs}

\subsection{Continuum Shape} 

In the {\it ASCA} sample, a power law component is detected in 100\%
of BLRGs and QSRs and in 90\% of NLRGs and RGs above 2 keV. The
average photon index is $\langle \Gamma_{2-10~keV} \rangle \sim
1.7-1.9$ for all the four subclasses, in agreement with unification
models, which postulate that both type-1 and type-2 AGNs sport the
same type of central engine, although view from a different direction.
Indeed, in NLRGs and RGs the nuclear power law spectrum is heavily
absorbed. The column densities detected by {\it ASCA} are $N_{\rm H}
\sim 10^{21-24}$ cm$^{-2}$ and are most likely due to the putative
obscuring torus on parsec-scales (e.g., Urry \& Padovani 1995). More
surprising is the detection of similar columns of {\it cold} gas in a
fraction of BLRGs and QSRs, which we discuss further below.

\begin{figure}
\noindent{\psfig{file=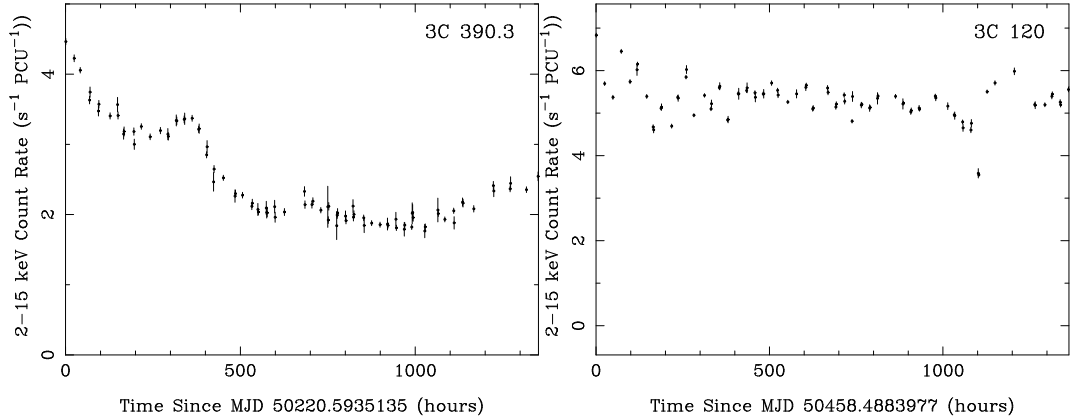,width=5.5cm,angle=-90}}{\psfig{file=3c120-lc.ps,width=5.5cm,angle=-90}}
\caption[]{Light curves from {\it RXTE} monitoring observations of the
BLRGs 3C~390.3 and 3C~120 in 1997 May and 1997 January,
respectively. The light curves are binned at 4600 s. Flux variability
is apparent, with different behaviors for the two sources. While
3C~390.3 shows a long-term trend, with large-amplitude variations on
timescales of \ltsima 2 weeks, ``flickering'' variations are observed
for 3C~120.}
\end{figure}

We compared the photon index distributions of the RL objects of our
sample to those of RQ AGNs of matching X-ray luminosity, and found
that the distributions of X-ray continuum slopes for the two classes
are not demonstrably different. This finding is consistent with
previous results that a flat X-ray component in RL AGNs is due to the
beamed contribution of the jet (see above). This conclusion is
bolstered by the fact that no correlation is found between the nuclear
X-ray and core radio luminosities. This was expected since the AGNs in
our sample are are associated with lobe-dominated radio sources.

\subsection{Continuum Flux and Spectral Variability} 

While X-ray variability on short-timescales (from hours to minutes) is
common in Seyfert 1s (e.g., Edelson 2000), no pronounced flux
variations are observed in the BLRGs and QSRs of our {\it ASCA} and
{\it RXTE} samples on timescales $<$ 10 hours. The only exception is
3C~120, where flux changes with amplitude $\sim$ 20\% are present in
the archival {\it ASCA} 40 ks observation.  The lack of pronounced
variability may be a consequence of the high luminosity of these
objects since Seyferts of comparable luminosities have similar
variability properties.

\begin{figure}
\centerline{\psfig{file=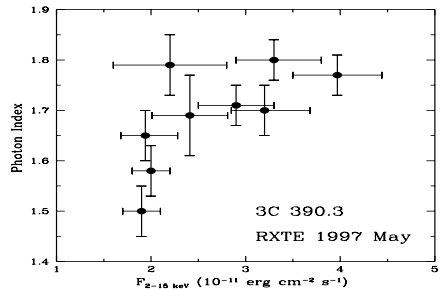,width=6cm,height=4cm}}
\caption[]{Spectral variability of 3C~390.3 during the {\it RXTE}
intensive monitoring in 1997 May (see Figure 1). The photon index from
fits to time-resolved {\it RXTE} spectra is plotted versus the
observed flux.  There is a trend of steeper slopes for increasing
flux, an a large and probably intrinsic dispersion in spectral indices
at lower fluxes.}
\end{figure}

In contrast, significant X-ray flux and spectral variations on longer
timescales are observed in two well-monitored BLRGs, 3C~390.3 and
3C~120. Figure 1 shows the {\it RXTE} 2--15 keV light curves obtained
during the intensive monitorings in 1996 May and 1997 January, with a
sampling characterized by short pointings roughly once a day for 60
days.  Flux variability is readily apparent in both cases, with
qualitatively different behaviors. 3C~390.3 exhibits a trend of
decreasing flux by a factor 2 in $\sim$ 20 days, with smaller changes
($\sim$ a few percent) on timescales of \ltsima 4 days
superposed. Non-linear variability was found in the soft X-ray light
curve of this object from previous {\it ROSAT HRI} observations
(Leighly \& O'Brien 1998). In contrast, 3C~120 exhibits more erratic
flux variations similar to ``flickering'', with small-amplitude
intra-day flares superposed on a constant baseline.

We investigated accompanying spectral variability by integrating PCA
spectra during selected intervals in the light curves in Figure 1
corresponding to high, intermediate, and low states. We fitted the
spectra in 4--20 keV with a model consisting of a power law, absorbed
by the Galactic absorbing column, plus the \feka\ line at 6.4
keV. This model was chosen because it describes well the average
PCA+HEXTE spectrum. Figure 2 shows the plot of the best-fitting photon
index versus the 2--10 keV flux in the case of 3C~390.3. Spectral
variability is readily apparent, with the spectrum becoming steeper as
the flux increases. Note, however, the large dispersion of slopes at
low flux levels (\ltsima 3 $\times 10^{-11}$ erg cm$^{-2}$ s$^{-1}$),
which is larger than the typical error bar, indicating a true
intrinsic dispersion of values.  A similar correlation between the
X-ray slope and flux was previously observed in 3C~120 (Halpern 1985)
and is consistent with the general trend of steeper-when-brighter
observed in Seyfert 1s (Grandi et al. 1992).

\subsection{The \feka\ Line} 

At high energies, the \feka\ line is detected with {\it ASCA} in 67\%
of BLRGs, 20\% of QSRs, 33\% of NLRGs, and 30\% of RGs. It is
generally broad in type-1 sources, with FWHM \ltsima
50,000~km~$^{-1}$s, and unresolved in NLRGs and RGs. However, in most
cases it is difficult to study the line profile due to the limited
sensitivity of {\it ASCA} and the fact the line is weak in RL AGNs.
More stringent constraints are provided by {\it RXTE} thanks to its
larger collecting area.  The \feka\ line is detected in the BLRGs of
the {\it RXTE} sample with an Equivalent Width of EW \ltsima 100 eV,
lower than what was measured by {\it ASCA} in most Seyferts (Figure
3a).  Unfortunately, because of the poor {\it RXTE} resolution, the
lines are unresolved, with {\it FWHM} \ltsima 50,000~km~s$^{-1}$ at
90\% confidence. 

\noindent{\bf 3C~390.3.} Interesting constraints on the \feka\  line
profile are obtained for the nearby, bright BLRG 3C~390.3.  
The \feka\ line profile is better fitted by a disk-line model than by
a Gaussian model, with EW=132$^{+40}_{-48}$ eV; however, the inner
disk radius is not constrained. A deep (134 ks) {\it RXTE} observation
shows that the line is unresolved, with width FWHM \ltsima
24,000~km~s$^{-1}$ (Figure 3c). These findings lend support to the
idea that the \feka\ line in RL AGNs comes from the cold, thin
accretion disk exterior to an ADAF. No reflection component is
required to fit the {\it RXTE} data above 10 keV, with an upper limit
to the covering factor $\Omega/2\pi$ \ltsima 0.5 at 90\% confidence.

\noindent{\bf 3C~382.}  We observed 3C~382 simultaneously with ASCA
and {\it RXTE} for 100 ks and 60 ks, respectively, in 1999 March, in
order to study in detail the profile of the \feka\ line. This source
was selected because it showed an extremely broad (FWHM $\sim$
170,000~km~s$^{-1}$) line in a 40 ks {\it ASCA} exposure. The
combination of spectra from this these two instruments lends itself to
studying the line profile because it provides wide-band coverage and
good resolution and collecting area in the energy range 6--7~keV. Thus
it allows a reliable determination of the continuum underlying the
emission line. 
\footnote[1]{Unfortunately, it was later discovered that there are 
calibration discrepancies between the two instruments, which hamper 
the determination of the continuum. In practice, during the joint {\it
ASCA+RXTE} fits the slope and normalization of the {\it ASCA} and {\it
RXTE} continua were left untied.}.

The 0.6--50 keV continuum is well described by a power law with photon
index $\Gamma=1.8-1.9$ and observed flux $F_{2-10~keV} \sim 5-6 \times
10^{-11}$~erg~cm$^{-2}$ s$^{-1}$, plus a bremsstrahlung plasma
component with $kT \sim 0.4$ keV below 1 keV. The power law flux is a
factor of $\sim$ 2 higher than during the previous {\it ASCA} and {\it
RXTE} observations.
The simultaneous {\it ASCA} and {\it RXTE} observations confirm that
the \feka\ line is quite broad compared to other BLRGs: we measure a a
width of FWHM 78,000$^{+31,000}_{-25,000}$~km~s$^{-1}$ at 90\%
confidence, and with EW=228$^{+131}_{81}$~eV, resolved at $>$ 99\%
confidence.  No improvement to the fit is obtained when a disk-line
profile is used, nor when the reflection component is added. The upper
limit to the solid angle subtended by any medium producing Compton
reflection $\Omega/2\pi$ \ltsima 0.5 at 90\% confidence.

The large width of the \feka\ line in 3C~382 is puzzling and requires
explanation. If the line is indeed as broad as we have measured it to
be, this source stands out as the only BLRG with a Seyfert-like \feka\
line profile. Alternatively, the line could be a blend of different
components originating in various parts of the accretion flow and/or
the jet. We tested this idea adding a narrow unresolved Gaussian to
the fit to the {\it ASCA+RXTE} data; while the fit is not improved,
the degraded SIS resolution at the time of these observations does not
allow us to rule out the presence of multiple components to the \feka\
line in 3C~382. Thus, the final solution to this puzzle will have to 
await observations at higher spectra resolution than {\it ASCA} can 
currently deliver.

\subsection{The Reflection Component} 

The wide energy range (2--250 keV) covered by the PCA and HEXTE
instruments on {\it RXTE} is well suited to the study of the Compton
reflection in RL AGNs at energies above 10~keV. The strength of this
component is parameterized in terms of $R=\Omega/2\pi$, i.e., the
fraction of solid angle subtended by the reprocessor to the
illuminating source. We find that the reflection component is
generally weak or undetected in the BLRGs observed with {\it RXTE},
with strengths $R$ \ltsima 0.4--0.5. This is much weaker than
what is typically observed in Seyfert 1s (Figure 3b). 

Interestingly, the spectrum of the NLRG Centaurus~A, includes an
unresolved \feka line and no reflection component (Rothschild et
al. 1999), consistent with unification models.

\section{Possible Interpretations} 

In the currently accepted accretion scenario for RQ AGNs, X-ray
emission is produced in a hot ``corona'' overlaying and illuminating a
standard, Shakura-Sunyaev accretion disk (e.g., Haardt, Maraschi, \&
Ghisellini 1994).  In this picture the reprocessor (disk) subtends a
solid angle of $\Omega=2\pi$ to the illuminating source, as observed
by {\it Ginga} and {\it RXTE} in Seyfert galaxies. The \feka\ line is
emitted within a few gravitational radii from the black hole (Fabian
et al. 1989), and has a skewed profile with a broad red wing.  Its
equivalent width depends on the solid angle subtended by the disk
to the X-ray source.

In BLRGs we observe weaker reflection components, indicating $\Omega
\approx \pi$. From this perspective, one possible scenario is that the
inner parts of the disk (below some transition radius) inflate under
the pressure of the ions which are much hotter than the electrons,
thus forming an ion torus or ADAF. The ion torus responsible for the
emitted continuum radiation from radio to IR via synchrotron,
bremsstrahlung, and/or inverse Compton emission. The cold, thin disk
exterior to the ADAF produces the reflection component and the \feka\
line; both of these features would be weaker than in Seyferts since
the disk subtends a solid angle $\Omega\approx \pi$ to the X-ray
source. 

For an assumed black hole mass of $M=10^8 M_{\odot}$, the lack of
X-ray flux variability on timescales \ltsima 0.4 days constrains the
light-crossing radius of the emitting region to be \gtsima 100 $R_{\rm
g}$ (where $R_{\rm g}\equiv GM/c^2$ is the gravitational radius, with
$M$ the mass of the black hole). This is consistent with the expected
light-crossing size of an ADAF, which is somewhere between a few
light-hours and a few light-days.  The ADAF scenario can also
accommodate X-ray slopes similar to Seyferts, depending on the
transition radius between the inner ADAF and the disk. 

\begin{figure}
\noindent{\psfig{file=mike_fig7.ps,width=6.5cm,angle=-90}}
{\psfig{file=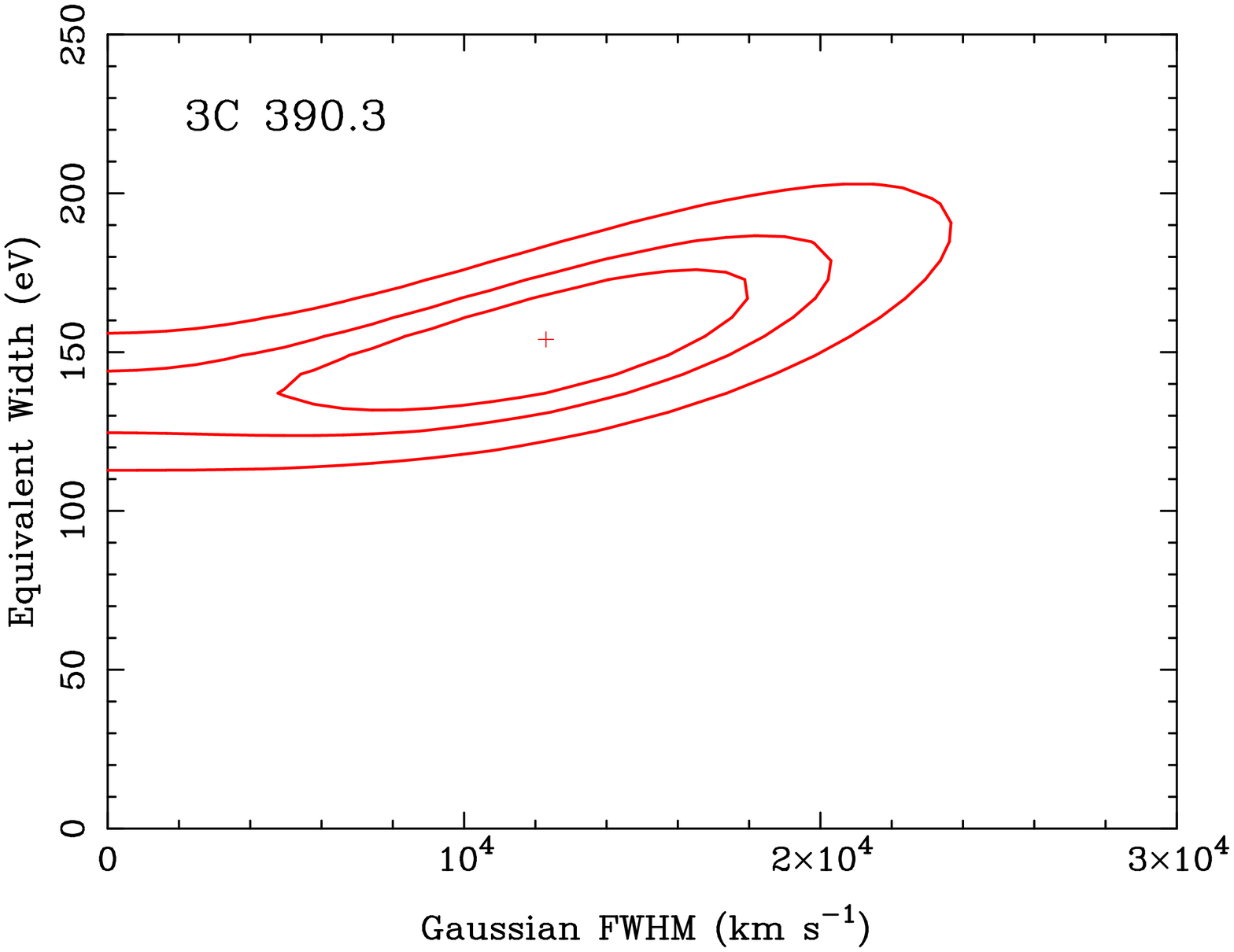,width=6cm,angle=-90}}
\caption[]{{\it (a) Left and (b) Center:} Histograms of the rest-frame
EW of the \feka\ line and covering fraction of the reflection
component, respectively, for the BLRGs of our {\it RXTE} survey
(shaded) and for Seyfert 1s studied by {\it ASCA} and GINGA (open),
from the literature (Nandra et al. 1997; Nandra \& Pounds
1994). Radio-loud AGNs have weaker lines and Compton humps than
radio-quiet AGNs.  {\it (c) Right:} Contours at 68\%, 90\% and 99\%
confidence for the EW of the \feka\ line versus its velocity width
from the fit to an archival 134 ks {\it RXTE} observation of
3C~390.3. The line is marginally resolved, with an interesting upper
limit to its width of FWHM \ltsima 24,000~km~s$^{-1}$.}
\end{figure}

An alternative scenario was proposed by Wo\'zniak et al. (1998) on the
basis of similar results to ours, obtained from non-simultaneous,
archival {\it ASCA}, {\it CGRO}/OSSE, and {\it Ginga} data. In that
scenario the central engines of BLRGs harbor a standard, geometrically
thin disk but the bulk of the X-ray continuum is produced by the inner
parts of a {\it mildly} relativistic jet. The X-rays from the jet are
emitted in a cone that is wide enough to illuminate the obscuring
torus but not wide enough to illuminate the accretion disk.  Thus,
these authors proposed that the Compton reflection component and the
\feka\ line are originate in the obscuring torus (the former via
Thomson scattering and the latter via fluorescence). However, our
result that RL AGN have similar X-ray slopes to RQ AGNs argues against
a beamed component in the X-rays. Moreover, the scenario suffers from
the additional drawback that X-rays from the jet are unlikely to be
beamed in a wide-angle cone: since jet Lorentz factors are thought to
have values around 10, the opening angle of the beam should be less
than $10^{\circ}$. Therefore, the obscuring torus will not be
illuminated by this beam.

These scenarios can in principle be tested further by studying the
profile of the \feka line in detail with instruments such as {\it XMM}
and {\it Astro-E}. If the ADAF scenario is correct, the line will be
double peaked with FWHM \ltsima 20,000~km~s$^{-1}$ (as indicated in
the case of 3C~390.3), corresponding to an inner radius of a few
hundred $R_{\rm g}$ or more. If the \feka\ line is produced in the
obscuring torus, i.e., at a very large distance from the central black
hole, then its profiles will be rather narrow (FWHM $\sim$
300~km~s$^{-1}$), and thus unresolved by the {\it XMM} detectors and
possibly even unresolved by the {\it Astro-E} calorimeter. Finally, if
the \feka\ lines of BLRGs have the same profiles as those of Seyfert 1
galaxies, then these will be easily measurable in high signal-to-noise
ratio spectra obtained with {\it XMM}.

It has also been suggested (Blandford \& Znajek 1977; Koide et
al. 1999) that RL AGNs harbor more rapidly spinning black holes than
RQ sources. While this mechanism would provide a ready explanation for
the formation and collimation of the jets, its implications for the
broad-band and X-ray properties of the two AGN classes are less
clear. Nevertheless, it should be possible at least in principle to
test this model through the shape of the \feka\ line (see Reynolds
2000 and references therein). This awaits the large collecting areas
and resolutions of future X-ray missions, such as the proposed {\it
Constellation-X} (NASA) and {\it XEUS} (ESA).

\section{X-ray absorption in RL AGNs} 

X-ray observations of RL AGNs also provide constraints on the gas
surrounding their central engines. While ionized absorption is common
in Seyfert 1s (Reynolds 1997; George et al. 1998), no analogous
evidence is found in the BLRGs of our sample, except in 3C~390.3.
Instead, large columns of {\it cold} gas are detected in 50\% of BLRGs
and QSRs. In some case the absorbing columns are as high as those
found in NLRGs and RGs. This is in contrast to unification models
where the line of sight to type-1 sources should be devoid of cold
material.

The columns detected in BLRGs and QSRs of our sample are similar to
the columns observed in more distant RL QSRs with {\it ASCA} and {\it
ROSAT} (Elvis et al. 1998; Cappi et al. 1997). This is apparent from
Figure 4, where the excess X-ray column $N_{\rm H^{exc}}$ (in the
source's rest-frame) is plotted versus the nuclear 2--10 keV
luminosity. While there is a large dispersion in the values of the
column density at lower luminosity, no trends are present over more
than five decades: the more distant sources have similar intrinsic
absorbing columns to the nearby sources. Interestingly, while the
BLRGs and QSRs of our sample are lobe-dominated, the more distant
objects are core-dominated. This suggests that the absorber, common to
both low- and high-redshift sources, must be isotropic in the sources'
rest-frame, or subtend a relatively large solid angle to the central
illuminating source. As discussed by Halpern (1997), the nature of the
absorbing medium and its location relative to the emission line
regions surrounding the central engine is unclear. It is plausible to
place the X-ray absorber between the broad-line region and the central
engine, as long as it does not contain a significant amount of
dust. In this context the absorber could be a wind outflowing from the
central engine (e.g., Blandford \& Begelman 1999).

\begin{figure}
\centerline{\psfig{file=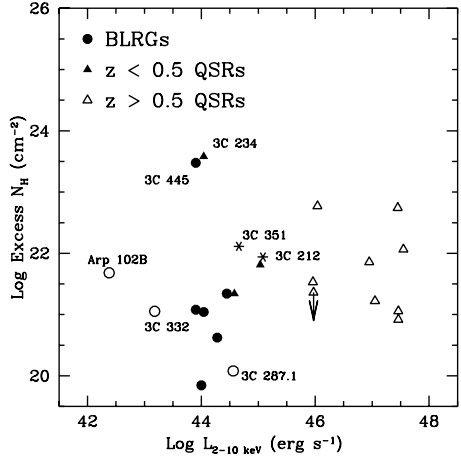,width=6cm}}
\caption[]{Plot of the intrinsic (rest-frame) excess $N_{\rm H}$
versus luminosity for the BLRGs and QSRs of our study, together with
more distant sources studied with {\it ROSAT} and {\it ASCA} in the
literature. No trends of the column density with luminosity are
apparent over more than five decades.}
\end{figure}

\section{Conclusions} 

Recent observations at high sensitivity and resolution with {\it ASCA}
and {\it RXTE} have shown that systematic differences in the X-ray
properties of RL and RQ AGNs. The  
X-ray results suggest that the origin of the RL/RQ AGN dichotomy must
be sought in the intrinsic properties of the central engines of these
systems. 

Much remains to be done to further our understanding of RL AGNs.  The
three major X-ray observatories of the next century, {\it Chandra},
{\it XMM}, and {\it Astro-E}, will have a central role as they will
allow us to study in detail the \feka\ line profile, the nature of the
mysterious X-ray absorber, and X-ray variability on the shortest
timescales.  Future X-ray observations at high sensitivity and
spectral resolution will provide a giant leap forward in our
understanding of RL AGNs (comparable to the one {\it ASCA} already
provided for RQ AGNs), opening a new perspective on the origin of the
RL/RQ AGN dichotomy. 

\begin{acknowledgements}
We acknowledge support from NASA contract NAS--38252 and NASA grants 
NAG5--7733 and NAG5--8369. 
\end{acknowledgements}

\end{document}